\documentstyle[aps,amsfonts,epsfig]{revtex}

\begin{document}

\title{Increasing Superconducting T$_c$'s by a Factor of 1000 \\ 
with StripeLike Hopping Anisotropies}
\author{Saurabh Basu, A. Callan-Jones\cite{byline}, and R. J. Gooding}
\address{Department of Physics, Queens University,  
Kingston, Ontario K7L 3N6, Canada} 

\date{\today}

\maketitle


\begin{abstract}

We have studied the enhancement of the superconducting transition temperature,
$T_c$, in a $t-J-U$ model of electrons moving on a square lattice in
which anisotropic electronic hopping is introduced. The inclusion
of such hopping mimics, in a approximate fashion, 
a potentially  important characteristic of materials possessing 
stripelike charge and spin correlations. For this model we have calculated 
$T_c$ for singlet pairing using the non self-consistent 
Thouless criterion, and find a dramatic enhancement of $T_c$ induced by 
hopping anisotropies. Further, the maximum increase in $T_c$  is obtained when 
the system is pushed towards the 
{\em extreme anisotropy limit}, that is, when the hopping of electrons
is confined to occur in $1+0^+$ dimensions. We demonstrate that in this
limit the increase in $T_c$, with respect to the isotropic system,
can be of the order of 1000. We have also determined that 
in the extreme anisotropy limit the superconducting gap is an equal mixture of 
$s$ and $d$ pairing symmetries (two choices of such a combination being 
$s + d$ and $s + id$) owing to the reduced (square to rectangular) symmetry 
of the system in the presence of hopping anisotropies. Thus, the presence of 
$d$-wave superconducting features in materials whose symmetry is very
different from that of a two-dimensional square lattice, with the
anisotropy produced by the appearance of stripes, is not unexpected.

\end{abstract}


\section{Introduction}

Strong support for the existence of stripes in high-$T_c$ cuprates,
and other transition-metal oxides, has been provided by
many experiments \cite{Tranquada}. This leads to the question:
Do stripe correlations help, hinder, or even possibly create
the pairing instability that leads to superconductivity?
In this paper we demonstrate that at least one feature of
stripelike correlations strongly enhances the superconducting transition
temperature, $T_c$.

A previous examination of the magnetic properties of the very weakly doped
cuprates \cite {cneto} modeled the observed experimental support
for stripe correlations \cite {borsa} using an effective Hamiltonian
in which a (spatially) anisotropic exchange interaction was implemented
to represent the stripe-induced magnetic energy scales. That is,
in the direction parallel to the stripes the full local $Cu-Cu$
exchange would be present, while perpendicular to the stripes
a reduced exchange would be encountered across such stripes.
Renormalized Hamiltonians of a similar simplifying spirit were also
used in other studies of the doped cuprates \cite {aharony,doniach,skyrms}.

In a recent paper we introduced a model that mimics one aspect of the stripe 
correlations by incorporating an anisotropic hopping Hamiltonian \cite
{sbrjgpwl}; that is, carriers are
expected to be able to move much more readily along the direction of the
stripes, so-called rivers of charge \cite {borsa}, than 
perpendicular to the stripes.
Allowing the carriers to interact via Heisenberg superexchange (using
a two-dimensional $t - J$ model), and excluding double occupancy ($U
\rightarrow \infty$), we have found exact solutions for the two electron
bound-state problem, {\em viz.}, in the dilute electron density limit. Our
calculations demonstrated how stripelike hopping anisotropies
can produce (i) a vanishing value for the threshold exchange coupling,
$J_c/t$, that is required for the stabilization of two-particle bound
states, and (ii) a dramatic increase in the two-electron binding energies
\cite{sbrjgpwl}. The physical interpretation of these results is
particularly interesting --- for the same problem in either one
dimension, or in two dimensions with isotropic hopping, a critical superexchange of
$J_c/t = 2$ is obtained. However, in the extreme anisotropy limit
the electrons are confined to hop in only one lattice direction,
and for such a system we found that an infinitesimal $J/t$ produces bound states. The
explanation of this result follows from recognizing that in the
bound state the two
electrons will travel in opposite directions on near-neighbour ``chains",
and when they are on near neighbour sites, and have anti-parallel
spins, a $S^+S^-$ spin exchange interaction flips the chains on which
the electrons are travelling. An infinitesimal $J/t$ using this
mechanism is sufficient to produce a two-electron pair. It is
appropriate to think of this result as binding produced by dimensional
confinement.

This interesting result naturally leaves us with the task
of investigating the transition to a superconducting phase in a system with a nonzero
electronic density as the system undergoes a crossover from a two-dimensional
to a 1+0$^{+}$-dimensional system induced by hopping anisotropies.

In this paper we present the results of our investigation on the
effects of such stripelike hopping anisotropies on 
$T_c$. We use the non self-consistent Thouless
criterion \cite {Thouless} to determine $T_c$, a procedure that is known to reproduce
the BCS result for the superconducting transition temperature \cite{ambegaokar69}. 
We find a dramatic enhancement of $T_c$
the greater the degree of hopping anisotropy, and the maximum $T_c$
obtained is found to saturate in the extreme anisotropy limit. For
certain system parameters, this enhancement can be of the order of 1000! 
Thus, this simple model of hopping anisotropy, which may be an essential 
ingredient to stripe correlations, provides a robust demonstration supporting 
the conjecture that stripes can indeed augment pairing correlations.

\section{Model Hamiltonian}

We model stripelike hopping anisotropy in a square lattice of strongly
interacting electrons using the following $t - J - U$ model:
\begin{equation}
\label{eq:tJU}
H= -\sum_{\langle i,j \rangle,\sigma}t_{ij}(c^{\dagger}_{i,\sigma}
c_{j,\sigma} + {\rm {h.c.}})
+{\sum_{\langle i,j \rangle}}J_{ij}({\bf {S_{i} \cdot S_{j}}}
-\frac{1}{4}n_{i}n_{j}) + U\sum_i n_{i,\uparrow}n_{i,\downarrow}~~.
\end{equation}
In this Hamiltonian, the sites of a two-dimensional square lattice of size
$L_x \times L_y$ with periodic boundary
conditions are labeled by the indices $i$ and $j$, $t_{ij}$ and $J_{ij}$
are the hopping integrals and exchange couplings between sites $i$ and
$j$, respectively, $c_{i,\sigma}$ is the annihilation operator for 
electrons at site $i$ of spin $\sigma$, 
$n_{i,\sigma}$ is the number operator for electrons at site $i$ with
spin $\sigma$,
and $U$ is the on-site Hubbard energy.

The most familiar strong-coupling variant of the Hubbard model is the
$t-J$ model, and the physics of (square lattice) doped Mott insulators
described by this model was reviewed by Dagotto \cite {elbio_rmp}. As
emphasized by, {\it e.g.}, Anderson \cite {pwa_97}, a vital component
of the $t-J$ Hamiltonian is the constraint of no double occupancy. That
is, in the $t-J$ model one does not use the electron creation and
annihilation operators of Eq.~(\ref{eq:tJU}), but rather one uses
constrained creation and annihilation operators (for example, see the
discussions in Ref.~\cite{elbio_rmp}). However, the above $t-J-U$
Hamiltonian can be used to accomplish this same mathematical projection
by taking $U\rightarrow \infty$ --- this simplifying approach has been
noted by a variety of researchers (see, {\it e.g.},
Refs.~\cite{Lin,Pethukov,kagan}), and will also be used by us.

In this paper 
we restrict $t_{ij}$ and $J_{ij}$ to be nonzero for near neighbours (NN) only.  
Further, as we did in Ref. \cite {sbrjgpwl}, we allow the hopping integral in the
$x$ direction, $t_x$, to be different than the hopping integral in the
$y$ direction, $t_y$. We have also investigated the physics that
arises when $J_x$ is allowed to be different than
$J_y$, but find that no qualitatively new physics arises as long as both
$J_x$ and $J_y$ remain nonzero. Thus, from now
on we set $J \equiv J_x = J_y$, and we analyze the resulting
hopping anisotropy problem in terms of
\begin{equation}
\label{eq:anist}
t_x \equiv t~~,~~~~~r = \frac{t_y}{t}~~.
\end{equation}
Thus, we have three dimensionless energy scales in the problem, 
{\em viz.}, $U/t$, $J/t$, and $r$; the $U/t \rightarrow \infty$ limit 
reduces this number to two.

\section{Identifying $T_{\lowercase{c}}$ in the Ladder Approximation}

We use the non self consistent Thouless criterion\cite{Thouless} to identify
the temperature at which our system becomes unstable with respect to a low temperature
superconducting phase. To this end, we determine the equation for the two-particle vertex,
or effective interaction, 
in the singlet channel. It is known that at low electron densities\cite{FW} one may evaluate
the vertex function via the ladder approximation to the Bethe-Salpeter equation, 
and the lack of convergence of the sum of the ladder diagrams identifies the critical 
temperature, {\it {viz.}} when the sum converges the system should be in the normal state. 
The non self consistent formulation of the ladder approximation allows us to work exclusively 
in the normal state, and has been shown to lead to an identical transition temperature to
that found in the BCS theory of superconductivity \cite {ambegaokar69}.
  
Thus, we focus on evaluating the effective interaction $\Gamma$ when all particle-particle 
diagrams are included. The integral equation for this function in this approximation can be written as 
(note that we only need to consider
$\Gamma_{\uparrow\uparrow,\downarrow\downarrow}$ \cite{ACJthesis}, and to 
eliminate the proliferation of spin indices in the equations that follow from now on we suppress 
all spin dependencies from our equations)
\begin{equation}
\label{eq:tmat}
\Gamma(k,k',Q) = V({\bf {k - k'}}) - \int \frac{d^{3}{\bf {q}}}
{(2\pi)^3}V({\bf {k-q}})G^{0}(q)G^{0}(Q-q)
\Gamma(q,k',Q)~~.
\end{equation}
The above integral equation is known as the Bethe-Salpeter equation 
in the ladder approximation \cite {FW}, and
the various functions appearing in this equation are defined as follows:  
$G^{0}(q)$ is the zeroth order single-particle Greens function defined
by,
\begin{equation}
G^0(q) = G^0({\bf q},i\omega_{n}) =  \frac{1}{i\omega_n
- (\varepsilon_{\bf q}-\mu)}~~;
\end{equation}
the fermionic Matsubara frequencies are given by $\omega_n =
(2n+1)\frac{\pi}{\beta}$;
for our anisotropic hopping model the single particle dispersion of
non-interacting band electrons is given by,
\begin{equation}
\varepsilon_{\bf q} = -2t(\cos q_x + r\cos q_y) ~~;
\end{equation}
$\mu$ is the chemical potential.

To proceed to the solution of the Bethe-Salpeter equation, we need to
reduce it according to symmetries. That is,
the anisotropy that is present in Eq.~(\ref{eq:tJU}) when $0\leq r <1$ reflects a 
lowering of the point group symmetry of the system from that of a square 
to that of a rectangle. Recall that the basis functions corresponding to the relevant
irreducible representations of the two-dimensional square lattice are 1, 
$(\cos k_{x} + \cos k_{y})$ and $(\cos k_{x} - \cos k_{y})$, which correspond 
to on-site $s$-wave, extended $s$-wave, and  $d$-wave gap symmetries,
respectively. Then note that in our anisotropic model these symmetries are
mixed and map onto the fully symmetric $A_1$ irreducible representation
of the rectangular point group. As a result, it is helpful to
decompose the interaction, in the singlet channel (even in ${\bf {k}}$), according to the
basis functions of $A_1$ of the rectangular point group symmetry. That is,
we artificially set $J_x \not= J_y$, analyze the resulting interaction
in terms of a linear combination of the $A_1$ basis functions of 1, $\cos k_{x}$, and  $\cos k_{y}$,
and then, in the last step of the calculation, restore the square lattice symmetry 
of the superexchange interaction by resetting $J_x = J_y$.

Following this prescription, and focussing on singlet pairing only,
we write the bare interaction term 
$V({\bf k - k'})$ as
\begin{equation}
\label{eq:vint}
V({\bf k-k'}) = U-2J_{x} \phi_x({\bf k})\phi_x({\bf k}^\prime) - 
2J_y\phi_{y}({\bf k})\phi_y({\bf k}^\prime)~~,
\end{equation}
where $\phi_x({\bf k}) = \cos {k_x}$ and $\phi_y({\bf k}) = \cos {k_y}$.
Substituting Eq.~(\ref{eq:vint}) into Eq.~(\ref{eq:tmat}) we obtain
\begin{eqnarray}
\label{eq:integ}
\Gamma(k,k',Q) & = &  [U - 2J_{x}\phi_x({\bf k})\phi_x
({\bf k}^\prime) - 2J_y\phi_y({\bf k})\phi_y({\bf k}^\prime)] \\ 
\nonumber
& - & \int \frac{d^{3}q}{(2\pi)^3} [U - 2J_{x}\phi_{x}({\bf k})
\phi_x ({\bf q}) - 2J_y\phi_y({\bf k})
\phi_y({\bf q})]
G^{0}(q) G^{0}(Q-q) \Gamma(q,k',Q) ~~.
\end{eqnarray}
The above equation is an integral equation for the vertex 
$\Gamma(q,k',Q)$, and demonstrates how the irreducible representations of
the square lattice group point are mixed
in the rectangular phase. To solve this equation we use a standard
projection technique ({\em e.g.}, see Ref. \cite{Lin}), and setting the total
centre-of-mass momentum ${\bf Q} = 0$, we obtain the following set of  equations:
\begin{equation}
\label{eq:detU}
\left [\begin{array}{lll}
(1+U\chi_0)  & ~-2J_{x}\chi_{x}  & ~-2J_{y}\chi_{y}  \\
~U\chi_{x} & (1-2J_{x}\chi_{xx}) & ~-J_{y}\chi_{xy} \\
~U\chi_{y} & ~-2J_{x}\chi_{xy}  & (1- 2J_{y}\chi_{yy}) 
\end{array} \right ] 
\left (\begin{array}{l} 
C_{0}({\bf {k}}) \\ C_{x}({\bf {k}}) \\ C_{y}({\bf {k}}) \\ 
\end{array} \right ) 
= \left [\begin{array}{lll} 
\chi_{0}  & ~-2J_{x}\chi_{x}\phi_{x}({\bf {k}})  & ~-2J_{y}\chi_{y}
\phi_{y}({\bf {k}})  \\ 
\chi_{x}  & (1 - 2J_{x}\chi_{xx}\phi_{x}({\bf {k}}))  & 
~-2J_{y}\chi_{xy}\phi_{y}({\bf {k}})  \\
\chi_{y}  & ~- 2J_{x}\chi_{xy}\phi_{x}({\bf {k}})  & 
(1 - 2J_{y}\chi_{yy}\phi_{y}({\bf {k}}))  
\end{array} \right ]
\end{equation}
We will also discuss results obtained in the no double occupancy limit, $U/t \rightarrow \infty$,
and in this limit Eq.~(\ref{eq:detU}) becomes 
\begin{equation}
\label{eq:det}
\left [\begin{array}{lll}
\chi_{0}  & ~-2J_{x}\chi_{x}  & ~-2J_{y}\chi_{y}  \\
\chi_{x} & (1-2J_{x}\chi_{xx}) & ~-2J_{y}\chi_{xy} \\
\chi_{y} & ~-2J_{x}\chi_{xy}  & (1- 2J_{y}\chi_{yy}) 
\end{array} \right ] 
\left (\begin{array}{l} 
C_{0}({\bf {k}}) \\ C_{x}({\bf {k}}) \\ C_{y}({\bf {k}}) \\ 
\end{array} \right ) 
= \left [\begin{array}{lll} 
\chi_{0}  & ~-2J_{x}\chi_{x}\phi_{x}({\bf {k}})  & ~-2J_{y}\chi_{y}
\phi_{y}({\bf {k}})  \\ 
\chi_{x}  & (1 - 2J_{x}\chi_{xx}\phi_{x}({\bf {k}}))  & 
~-2J_{y}\chi_{xy}\phi_{y}({\bf {k}})  \\
\chi_{y}  & ~- 2J_{x}\chi_{xy}\phi_{x}({\bf {k}})  & 
(1 - 2J_{y}\chi_{yy}\phi_{y}({\bf {k}}))  
\end{array} \right ]
\end{equation}
The various functions appearing in the above two equations are defined as 
\begin{eqnarray}
\label{eq:coeff} 
C_{0}  &=& \frac{1}{N}\sum_{{\bf {q}}}G^{0}({\bf {q}})
G^{0}({\bf {-q}})\Gamma({\bf {q,k}})~~,  \nonumber \\
C_{x} &=&   \frac{1}{N}\sum_{{\bf {q}}}\phi_{x}(q)
G^{0}({\bf {q}})G^{0}({\bf {-q}})\Gamma({\bf {q,k}})~~, \\
C_{y} &=&    \frac{1}{N}\sum_{{\bf {q}}}\phi_{y}(q)
G^{0}({\bf {q}})G^{0}({\bf {-q}})\Gamma({\bf {q,k}})~~. \nonumber
\end{eqnarray}
\newpage\noindent
The various ${\bf Q = 0}$ susceptibilities are functions of
$\mu$ and $T$, and are defined by
\begin{eqnarray}
\label{eq:chi} 
\chi_{0} &=& \frac{1}{N}\sum_k G^0(k)G^0(-k)~~, \nonumber \\ 
\chi_{x} &=& \frac{1}{N}\sum_k \phi_x({\bf k}) G^0(k)G^0(-k)~~, \nonumber \\
\chi_{y} &=& \frac{1}{N}\sum_k \phi_y({\bf k}) G^0(k)G^0(-k)~~, \\
\chi_{xx} &=& \frac{1}{N}\sum_k \phi_x({\bf k}) G^0(k)G^0(-k)~~, \nonumber \\ 
\chi_{xy} &=&   \frac{1}{N}\sum_k \phi_x({\bf k}) G^0(k)G^0(-k)~~, \nonumber \\
\chi_{yy} &=&   \frac{1}{N}\sum_k \phi_y({\bf k}) G^0(k)G^0(-k)~~. \nonumber
\end{eqnarray}

To study the pairing instability we
determine the temperature at which a divergence 
of the vertex $\Gamma(q,k',Q)$ occurs. Clearly, $\Gamma(q,k',Q)$ 
depends upon the coefficients $C_{0}$, $C_{x}$ and $C_{y}$, 
defined by Eq.~(\ref{eq:coeff}), and a singularity in them demands
the vanishing of the determinant of the coefficient matrix appearing on the 
left hand side of Eq.~(\ref{eq:det}); the vanishing of this determinant
is thus a simple way in which we can identify the superconducting transition
temperature, and hence $T_c$ is obtained numerically
as a function of the chemical potential $\mu$.
However, a different and perhaps more physical way of displaying our results is to 
determine $T_c$ as a function of the electron density per lattice site, $n$, and in the BCS
approximation $n$ at $T_c$ is given by 
\begin{equation}
n(\mu,T_c) = 2~\sum_{\bf k}\frac{1}{e^{-\beta_c(\varepsilon_{\bf k} -\mu)} + 1}~~.
\end{equation}
Below we use this latter equation to plot $T_c$ {\em vs.} $n$.

\begin{figure}
\begin{center}
\epsfig{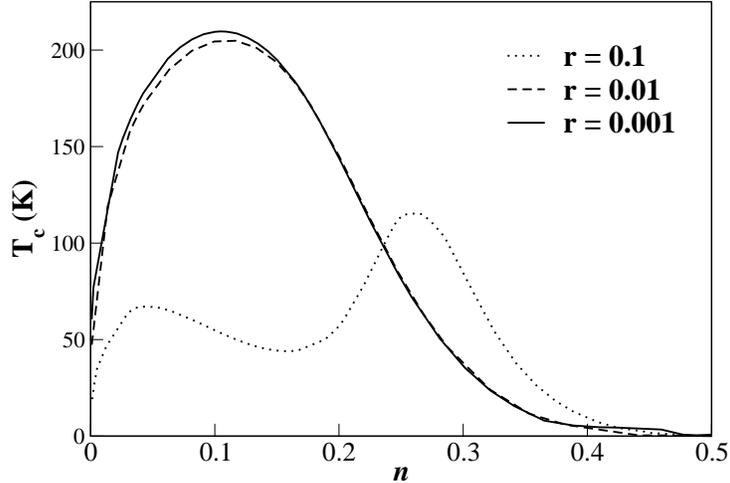}
\caption{The dependence of the superconducting transition temperature,
$T_c$, {\em vs.} the electronic density per lattice site, $n$, for 
$U/t = 0$ and $J/t = 1/3$. We have set $t = 1~eV$ and express $T_c$ in
Kelvin. The three curves correspond to anisotropic hopping ratios, $r$, of
$r = 0.1, 0.01,$ and 0.001.}
\label{fig:Ueq0_Tcvsr}
\end{center}
\end{figure}

Figure~\ref{fig:Ueq0_Tcvsr} shows our results for $U/t=0$ and $J/t=1/3$ (similar results
are found for other ratios of $J/t$) with a hopping anisotropy of $r~=~0.1,~0.01,$ and $0.001$. 
(We have set $t = 1~eV$ as a representative energy; also, this allows us to
express $T_c$ in Kelvin.)  As is known from the results corresponding to the isotropic case
\cite{Micnas,Feder}, for $U/t =0$ the low density ($\mu \sim -4t$)
region is dominated by (on-site) $s$-wave pairing, that is to say that
the maximum $T_c$ is obtained for $s$-wave pairing, whereas near
half filling ($\mu \sim 0$), the $d$-wave instability dominates. As $r$ deviates from
the isotropic case, the two pairing symmetries mix and we find a
decrease in the values of $T_c$ near half filling, and an enhancement 
in $T_c$ at low densities. For $r =0.1$
we see that there still exists a signature of two broad transitions 
($d$-wave and on-site $s$-wave). As $r$ decreases further the two transitions merge
with a $T^{max}_c \sim 200 K$.

\begin{figure}
\begin{center}
\epsfig{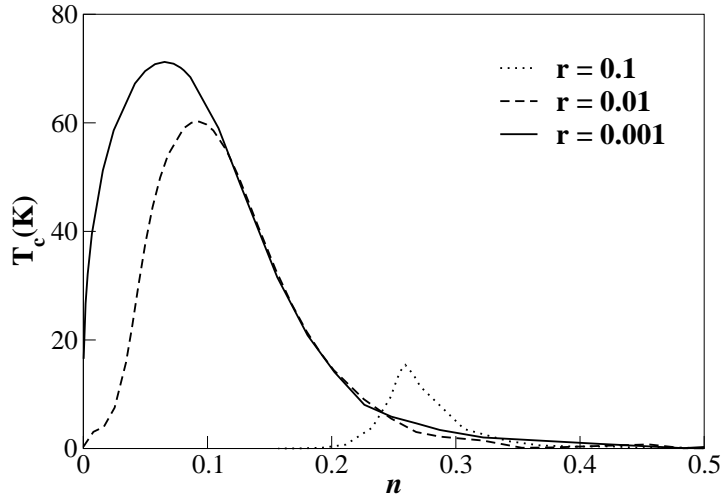}
\caption{The same as Fig.~\protect{\ref{fig:Ueq0_Tcvsr}}, where
the limit $U/t \rightarrow \infty$ has been implemented.}
\label{fig:Ueqinf_Tcvsr}
\end{center}
\end{figure}

Intermediate ratios of $U/t$ lead to results that smoothly interpolate between
$U/t=0$ and $U/t=\infty$, and in the interests of brevity we only show these
two limiting values of $U/t$. $U/t = \infty$ excludes the possibility of an on-site 
$s$-wave order parameter, hence the pairing symmetry has predominantly extended $s$- 
and $d$-wave components (see the discussion and numerical results in the next section). Our
$T_c$ {\it vs.} $n$ results for these parameters are shown in Fig.~\ref{fig:Ueqinf_Tcvsr}
for the same hopping anisotropies as in Fig.~\ref{fig:Ueq0_Tcvsr}. 

In both of these figures, it should be noted that when we take the
hopping anisotropy to
be even smaller than $r=0.001$, no noticeable change occurs --- thus,
our numerical
results for this ratio are representative of the $r\rightarrow 0$
extreme hopping anisotropy
limit.

The maximum $T_c$ at low densities, {\em viz.} $T^{max}_c \sim 70 K$
(again taking $t=1eV$)
occurs at about $n=0.08$. As a striking demonstration of the
effectiveness of stripelike
hopping anisotropy in increasing $T_c$, note for isotropic hopping,
$J/t=1/3$, $U/t=\infty$,
and a density of $n=0.08$ electrons per site,
we find that $T_c \sim 0.08 K$. That is, {\em we find an enhancement
of~} $T_c$ {\em due to hopping
anisotropy of a factor of about 1000!} Clearly, this dramatic increase
in the superconducting
transition temperature supports the conjecture that stripelike
correlations can strongly
affect pairing, and, at the very least, can greatly augment the
stability of the
superconducting phase at higher temperatures.

We note that results demonstrating an enhancement of $T_c$ due to orthorhombic distortions 
have been given earlier by  Li {\it {et al.}}\cite{Joynt1}, although the extreme anisotropy
limit was not considered; the importance of this latter limit was made apparent to us in
our study of the bound-state formation for the same Hamiltonian as in this paper, albeit for
only two electrons\cite{sbrjgpwl}.

We also note that the onset of superconducting pairing caused by interchain single particle
tunnelling (IST) has been considered earlier by Bourbonnais and Caron
\cite{Bourbonnais}. Starting from a Luttinger Liquid model (a linear
array of conducting chains separated by some distance), it was shown in the
limit of small interchain hopping IST leads to an effective
pair tunnelling which may eventually induce superconductivity in the
singlet channel below a temperature $T_{x^1}$ (signifying a crossover to a
higher dimensionality, {\em {viz.}}, a Luttinger Liquid to Fermi Liquid 
transition).

\begin{figure}
\begin{center}
\epsfig{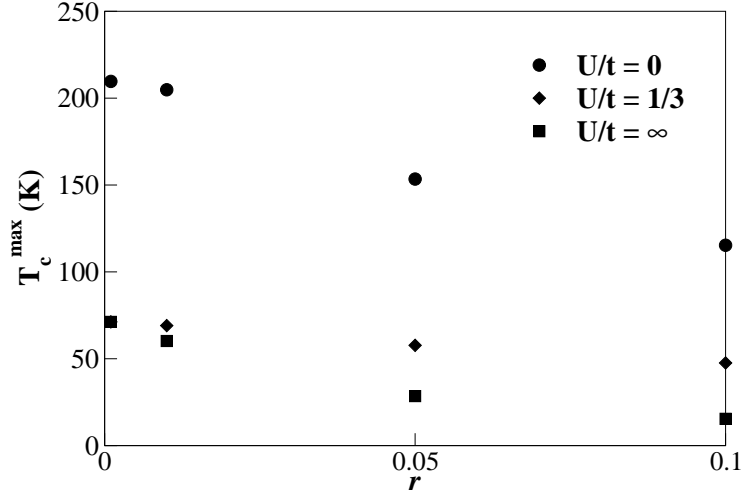}
\caption{
The maximum $T_c$, found as a function of electronic density for each
hopping anisotropy, is plotted as a function of the
hopping anisotropy $r$, for $U/t=0, 1/3$ and $\infty$ and $J/t =1/3$.
As in Figs.  1 and 2, we have expressed $T_c$ in Kelvin using $t = 1~eV$.}
\label{fig:Tcvsn}
\end{center}
\end{figure}

In the quasi one-dimensional systems studied in Ref.~\cite{Bourbonnais},
it was also found the transition temperature,
$T_c$, approximately scaled as $t_{\perp}$, the interchain hopping
integral, (or as $(t_{\perp})^{\alpha}$ where
$\alpha (\ge 1)$ is a continuous function of the interaction parameter). 
Our Fig.~\ref{fig:Tcvsn} shows the variation of the maximum $T_c$ (as a function
of electron density $n$) {\it vs.} hopping anisotropy, $r$, for 
$U/t =0, 1/3$ and $\infty$, and  for $J/t = 1/3$. The dependence is
found to be roughly linear, thus providing some support for the conjecture
made by Bourbonnais {\it {et al.}}\cite{Bourbonnais}. 

\section{BCS Gap Equation}

As mentioned earlier, anisotropic hopping couples the superconducting 
gaps in the $s$- and $d$-wave channels, and this leads one to think 
in terms of a mixing between the two symmetries having the following forms
\cite{Joynt1,Kotliar,Joynt2}: (i) an $s + id$ state, where there is a 
phase difference of $\pi /2$ between the gap functions in $s$- and $d$-channels,
and (ii) an $s + d$ state, where the phase difference is zero. 

We have evaluated
the ratio of these gaps by solving the zero temperature BCS gap
equation, which can be written as,
\begin{equation}
\label{eq:BCS}
\Delta ({\bf {k}}) = - \sum_{{\bf {k'}}} V({\bf {k,k'}})
\frac{\Delta({\bf {k'}})}{2E_{{\bf {k'}}}}~~,
\end{equation}
where $E_{{\bf {k}}} = \sqrt{(\varepsilon_{\bf {k}}-\mu)^2 + 
|\Delta ({\bf {k}})|^2}$. Note that
in the $U/t \rightarrow \infty$ limit, the on-site $s$-wave component is suppressed
and one is left to consider only the extended $s$- and the $d$-wave
symmetries. For simplicity, in this section we only report on 
this region of parameter space ($U/t \rightarrow \infty$).
This allows us to write the trial gap function in 
${\bf {k}}$-space with arbitrary phase difference $\theta$ in the form,
\begin{equation}
\label{eq:mix}
\Delta({\bf {k}}) = \Delta_{s} f_{s}({\bf {k}}) + e^{i\theta} \Delta_{d}
f_{d}({\bf {k}})~~,
\end{equation}
with $\Delta_{s}$ and $\Delta_{d}$ representing the amplitudes of the
gap functions, which are real, and 
\begin{eqnarray}
\label{eq:OPsyms}
f_{s}({\bf {k}}) &  = &  \cos k_{x} + \cos k_{y}~~, \\ \nonumber
f_{d}({\bf {k}}) &  = &  \cos k_{x} - \cos k_{y}~~. 
\end{eqnarray}
Ignoring the on-site term, we write the interaction in terms of the above-defined basis functions:
\begin{eqnarray}
\label{eq:VforDeltas}
V({\bf {k,k'}}) & = &  V({\bf {k - k'}}) = -2J[\cos (k_{x} -k'_{x}) + 
\cos (k_{y} -k'_{y})]  \\ \nonumber
& = & -J[f_{s}({\bf {k}})f_{s}({\bf {k'}}) + f_{d}({\bf {k}})
f_{d}({\bf {k'}})]~~.  
\end{eqnarray}
Using Eqs.~(\ref{eq:mix}-\ref{eq:VforDeltas}) in Eq.~(\ref{eq:BCS}), 
one can take the real and imaginary parts of Eq.~(\ref{eq:BCS}), 
and thus obtain coupled equations for $\Delta_{s}$ and  $\Delta_{d}$.
We have solved these equations numerically for both choices of the
relative phase $\theta$, {\it {viz.}} $\theta = \pi /2$ (corresponding to
$s + id$) and $\theta = 0$ (corresponding to $s + d$).
 
\begin{figure}
\begin{center}
\epsfig{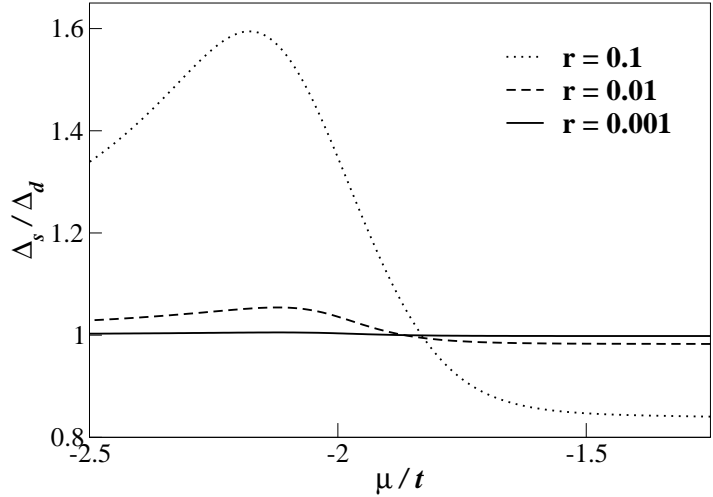}
\caption{The ratio of the gap amplitudes, {\em viz.}
$\Delta_{s}/\Delta_{d}$,
{\em vs.} the chemical potential (in units of $t$), for an $s+id$ order
parameter with $r=0.1, 0.01$ and $0.001$. The other parameters are
chosen to be $U/t=\infty$ and $J/t =1/3$.}
\label{fig:del_ratio_pi}
\end{center}
\end{figure}

Our numerical results for the ratio of gap amplitudes for $s+id$ are shown in
Fig.~\ref{fig:del_ratio_pi} --- since the results for $s+d$ are qualitatively
very similar, for brevity we omit that plot.
The $r=0.1$ case corresponding to the $s + d$ symmetry
shows a variation in the ratio $\Delta_{s}/\Delta_{d}$ from $\sim 0.8$ ($\sim
0.6$ for $ s +d$) for electronic densities near half filling, to $\sim 1.3$ 
($\sim 1.3$ for $s +d$) for
$\mu \simeq -2t(1+r)$ (low densities). Thus, at low densities the extended
$s$-wave contribution is somewhat larger, whereas
near half filling the $d$-wave component is stronger. 
As the anisotropy is increased, the ratios $\Delta_{s}/\Delta_{d}$ for
both $s+id$ and $s+d$ become almost equal to unity for all densities.

\vspace {0.5 truecm}

We note that similar results have been obtained earlier using variational 
Monte Carlo studies using a $t - J$ model \cite{Joynt1} where the 
authors found no distinguishable difference between $s +d$ and $s+id$ pairing symmetries.
More generally, they found that this result was true for any arbitrary $\theta$
in Eq.~(\ref{eq:BCS}). 

At this stage we cannot ascertain
which of the above pairing states is chosen by the system in presence of
hopping anisotropy, so there is a necessity to calculate the condensation
energies for these pairing symmetries using a microscopically derived 
Ginzburg-Landau free energy functional --- we leave this problem for a future publication.

\section{Conclusions}

To conclude, we find a dramatic enhancement of the superconducting transition 
temperature, $T_c$ at low electronic densities due to stripelike hopping anisotropies. 
Also, we find that the maximum $T_c$ is found to saturate as $r
\rightarrow 0$, that is, in the extreme anisotropy limit. In fact, 
one can show that $T_c$ tracks the electronic
density of states, consistent with the fact that our determination of
the superconducting instability is essentially a mean-field theory.
As the hopping anisotropy mixes different pairing symmetries, such as
on-site $s$-wave, extended $s$-wave, and $d$-wave, it is expected that the
system may choose either a $s +d$, or $s+id$, or more
generally a pairing of the form $\Delta_{s}({\bf {k}}) + e^{i\theta}\Delta_{d}
({\bf {k}})$, where $\theta$ may be a function of temperature and anisotropy.
To shed some light on this issue we have solved the zero temperature BCS gap 
equation and have evaluated the ratio between the gap amplitudes. Our 
results demsontrate an equal mixing between the $s$ and $d$ gap functions
in the extreme anisotropy limit.

We do not wish to give the impression that we believe that this model is
an adequate way of modelling the effects that stripelike rivers of charge
create. For example, we have ignored all features associated with the
discreteness of the stripes. Also, although we are including a near-neighbour
antiferromagnetic exchange, the ladder approximation that we are using
to determine a pairing instability will not be adequate near half
filling, and thus our theory is not just for a moderately doped antiferromagnet.

Nonetheless, the results that we published in Ref.~\cite{sbrjgpwl}
on two-electron bound state formation, as well as the substantial increase 
in $T_c$ that we are presenting in this paper, demonstrate the possible 
importance of including stripelike hopping anisotropies in realistic microscopic theories.

\acknowledgements
We wish to thank Claude Bourbonnais for directing our attention to a number of
helpful references. This work was supported in part by the NSERC of Canada.

\end{document}